\documentclass[11pt]{article}

\usepackage[utf8]{inputenc}
\usepackage[T1]{fontenc}
\usepackage{lmodern}
\usepackage{amsmath,amssymb}
\usepackage{booktabs}
\usepackage{longtable}
\usepackage{array}
\usepackage{calc}
\usepackage{enumitem}
\usepackage{microtype}
\usepackage{xcolor}
\usepackage{xurl}
\usepackage[hidelinks]{hyperref}
\usepackage[margin=0.9in]{geometry}
\usepackage{setspace}
\usepackage{etoolbox}

\providecommand{\tightlist}{%
  \setlength{\itemsep}{0pt}\setlength{\parskip}{0pt}}

\DeclareUnicodeCharacter{00D7}{\ensuremath{\times}}
\DeclareUnicodeCharacter{2265}{\ensuremath{\geq}}
\DeclareUnicodeCharacter{2264}{\ensuremath{\leq}}
\DeclareUnicodeCharacter{2260}{\ensuremath{\neq}}
\DeclareUnicodeCharacter{2248}{\ensuremath{\approx}}
\DeclareUnicodeCharacter{03BA}{\ensuremath{\kappa}}
\DeclareUnicodeCharacter{2013}{--}
\DeclareUnicodeCharacter{2014}{---}
\DeclareUnicodeCharacter{2018}{`}
\DeclareUnicodeCharacter{2019}{'}
\DeclareUnicodeCharacter{201C}{``}
\DeclareUnicodeCharacter{201D}{''}
\DeclareUnicodeCharacter{2212}{-}
\DeclareUnicodeCharacter{00A0}{~}

\hypersetup{colorlinks=true,linkcolor=black,citecolor=black,urlcolor=blue!50!black,pdftitle={The Custody Envelope Threshold},pdfauthor={Amadeus Brandes}}
\setlist[itemize]{leftmargin=1.25em,itemsep=0.15em,topsep=0.2em}
\setlist[enumerate]{leftmargin=1.5em,itemsep=0.15em,topsep=0.2em}

\title{\textbf{The Custody Envelope Threshold: Authority-Scaled Admission of External Artifacts in Institutional Infrastructure}}
\author{Amadeus Brandes\\
Independent Researcher, Germany\\
\texttt{brandesamadeus@gmail.com}}
\date{June 2026}

\begin{document}
\maketitle
\thispagestyle{plain}
\begin{abstract}
Modern infrastructure depends on externally maintained artifacts: package-registry dependencies, container images, CI/CD actions, Terraform providers, infrastructure modules, developer extensions, model weights, datasets, and emerging AI tool servers. These artifacts are easy for developers to fetch and compose, but difficult for institutions to admit, govern, revoke, or defend after failure. Existing software-supply-chain frameworks largely address producer integrity, secure consumption practices, maturity levels, or attack taxonomies. They do not explain why comparable artifact classes produce divergent institutional outcomes: some are admitted, some proxied, some policy-mediated, some vendor-mediated, some internalized, and some quarantined or rejected.

This article proposes the \textbf{Custody Envelope Threshold}: an authority-scaled model of artifact admission. The model has a normative core and a positive prediction. Normatively, an externally maintained artifact is defensible for direct institutional admission only when object identity, ingress path, and revocation capacity are sufficiently closed relative to the execution authority the artifact receives. Positively, the article argues that institutions exposed to repeated audit, incident-response, procurement, customer-assurance, and operational scrutiny tend to converge toward this authority-scaled custody rule over time. Deviations persist, but they appear as governance debt: tolerated in low-scrutiny or immature settings, then proxied, mediated, internalized, quarantined, or rejected when scrutiny rises.

The framework is operationalized as a four-condition ordinal instrument: object identity, ingress path, execution authority, and revocation capacity. Its central proposition is that custody requirements scale with delegated authority: a low-authority artifact may be admitted with modest custody controls, while a high-authority artifact requires strong identity, controlled ingress, and fast revocation before it can cross an institutional boundary. The article then adds a governance-selection heuristic grounded in transaction cost economics: when direct custody closure fails, the institution selects the lowest-burden governance mode that appears capable of closing the binding deficit at an acceptable risk/value tradeoff, where burden is shaped by delegated authority, local custody specificity, and the availability of external closure providers.

The framework is positioned against SLSA, NIST SSDF, Microsoft's Secure Supply Chain Consumption Framework (S2C2F), and existing software-supply-chain attack scholarship. S2C2F is the closest prior practical framework: it is explicitly consumption-focused, contains solution-agnostic practices organized into maturity levels, and emphasizes controlled OSS ingestion. The Custody Envelope Threshold differs by offering a descriptive-predictive model across heterogeneous artifact classes rather than a maturity/practice model for secure OSS consumption. SLSA and NIST SSDF provide integrity and secure-development evidence surfaces; supply-chain attack literature taxonomizes threats and safeguards. This article instead taxonomizes institutional admission responses and explains why different artifact classes bind on different custody conditions.

The theoretical lineage is the reference-monitor tradition, least privilege, and transaction cost economics. The reference-monitor and least-privilege traditions explain why authority-bearing artifacts require mediated admission; transaction cost economics explains why institutions choose different governance modes when direct admission fails. The article applies the model to six derivation cases - package dependencies, GitHub Actions, container images, Terraform providers and modules, VS Code extensions, and open model artifacts - then stress-tests it against \texttt{curl\ \textbar{}\ bash} installation flows and evaluates MCP servers/tools as a held-out artifact class. MCP is a strong held-out case because its servers expose tools that models can invoke, and current public implementations are already producing the governance surfaces the model predicts: registries, curated catalogs, gateways, scoped authorization, DLP-aware portals, and audit logs.

The central claim is falsifiable. Where authority is high and custody closure is weak, high-scrutiny institutions should not normalize direct admission; they should proxy, mediate, internalize, quarantine, or reject. Where strong custody closure exists relative to authority, admission should become more defensible even for externally maintained artifacts. If high-authority artifacts with weak identity, uncontrolled ingress, and weak revocation are routinely admitted into serious infrastructure without mediation under sustained scrutiny, the framework is weakened.
\end{abstract}

\section{1. The Empirical Puzzle}\label{the-empirical-puzzle}

Several externally maintained artifact classes face the same structural pressure. They are trivial for developers to pull into local work, build systems, CI/CD pipelines, runtime clusters, infrastructure automation, AI workflows, or developer environments. Yet their institutional outcomes diverge sharply.

Public package-registry dependencies are rarely banned outright; they commonly become routed through internal feeds, remote repositories, lockfiles, provenance checks, and approval workflows. Third-party GitHub Actions are often treated more harshly: the same institution that permits thousands of npm packages may block an unpinned marketplace action from touching CI secrets. Docker images split into trust strata: an arbitrary Docker Hub image, a Docker Official Image, a verified-publisher image, a hardened image, and an internally rebuilt base image do not have the same admission status. Terraform providers can become ordinary infrastructure dependencies because the ecosystem supplies signatures, lockfiles, checksums, and mirrors; public Terraform modules often get copied, wrapped, or republished internally because they encode architectural decisions rather than merely reusable code. VS Code extensions may be tolerated by individual developers but restricted through allowlists or private marketplaces at enterprise scale. Hugging Face model artifacts are useful enough to transform AI development, yet raw model pulls commonly remain better suited to experimentation when production use moves toward scanning, catalog mediation, managed deployment, or an enterprise hub.

The common pressure is external maintenance. The divergent outcome is institutional admission.

The procurement-adjacent question is therefore not ``is this artifact useful?'' or ``is this artifact open source?'' or even ``is this artifact secure?'' The question is narrower and more operational:

\textbf{What determines whether an externally maintained software artifact can be defended as admissible inside institutional infrastructure?}

The answer proposed here is the \textbf{Custody Envelope Threshold}.

An externally maintained artifact becomes institutionally admissible when the institution can close a custody envelope around the artifact before it crosses a meaningful execution boundary. The envelope is closed when three custody conditions - object identity, ingress path, and revocation capacity - are bounded strongly enough relative to the execution authority the artifact receives. If the envelope cannot be closed directly, the artifact is not necessarily rejected. It is proxied, mirrored, policy-mediated, vendor-mediated, quarantined, or internalized until the missing boundary is supplied by some other governance mode.

The central theoretical move is authority scaling. The required custody strength is not constant across artifacts. It rises with the authority delegated to the artifact. A documentation theme, a transitive logging library, a CI action with repository secrets, a Terraform provider with cloud API permissions, a base image for production workloads, an editor extension with source-code access, and a model artifact requiring remote code cannot be governed by the same admission standard.

The framework's core proposition is:

\textbf{Authority-Scaled Custody Closure.} For an artifact in a defined use context, direct institutional admission is defensible only when object identity, ingress path, and revocation capacity are each at least as strong as the execution authority delegated to the artifact. Where one or more custody dimensions fall below the authority level, the institution must either strengthen the deficient dimension, reduce the artifact's authority, substitute a lower-cost governance mode, or quarantine/reject the artifact.

This proposition is specific enough to be wrong. If high-authority artifacts with weak identity, uncontrolled ingress, and slow revocation are routinely admitted into serious production environments without mediation, the framework is weakened. If artifacts with strong identity, controlled ingress, bounded authority, and rapid revocation are still rejected for reasons internal to technical-institutional custody rather than separate legal, semantic, safety, or export-control gates, the framework is also weakened.

\section{2. Why the Standard Framings Fail}\label{why-the-standard-framings-fail}

The usual explanations are directionally useful and analytically insufficient.

``Software supply-chain security'' correctly names the risk category but cannot explain divergent institutional outcomes. npm packages, GitHub Actions, container images, Terraform providers, IDE extensions, and model weights can all introduce malicious or defective behavior. All can be maintained externally. All can be fetched through developer-friendly tooling. Security risk alone predicts generalized caution; it does not explain why one artifact class becomes proxied, another policy-mediated, another vendor-mediated, and another internalized.

``Developer convenience'' explains adoption pressure but not institutional permission. Developers pull public artifacts because doing so reduces time-to-use. Institutions admit artifacts only when platform engineering, security, procurement, compliance, and operations can defend the admission decision after an audit finding, outage, vulnerability, credential exposure, maintainer compromise, or release failure.

``Open source popularity'' explains demand but not admissibility. A package with millions of downloads may still be inadmissible by direct public fetch. A less popular provider may be admissible if it arrives through a signed registry, pinned version, internal mirror, checksum, and controlled execution path. Popularity increases pressure to admit; it does not supply the admission record.

Provenance, SBOMs, signing, and secure-development frameworks matter, but they are evidence surfaces rather than the admission model itself. SLSA addresses standards and controls for preventing tampering and improving supply-chain integrity. NIST SSDF provides secure software-development practices and a vocabulary for producers, acquirers, and operators. Those frameworks improve the evidence available to the admitting institution. They do not, by themselves, answer whether a specific third-party artifact may enter a CI runner with production credentials, a Kubernetes cluster, a developer laptop, an infrastructure state workflow, or an AI serving path.

The missing object is admission.

\section{3. Scope and Definitions}\label{scope-and-definitions}

The Custody Envelope Threshold applies to externally maintained artifacts that an institution pulls, installs, executes, composes, deploys, or serves inside infrastructure it governs. The relevant artifact classes include package-registry dependencies, CI/CD actions, container images, base images, Terraform providers, Terraform modules, Helm charts, Kubernetes operators, developer-environment extensions, model weights, datasets, and AI infrastructure components.

The framework does not apply cleanly to SaaS procurement, where the institution consumes a service rather than admitting a discrete artifact into its own execution environment. It is not a framework for commercial open-source license changes, fork dynamics, or vendor capture after dependency formation. It is not a general model of AI safety, legal suitability, export control, privacy compliance, or model behavior. Those concerns may block use after technical custody is closed, but they are separate downstream gates. The framework concerns a narrower institutional act: allowing an externally maintained object to become part of a controlled technical system.

Four terms are used precisely.

\textbf{Artifact} means a discrete externally maintained object that can be fetched, installed, executed, imported, deployed, loaded, or served.

\textbf{Admission} means institutional permission for an artifact to cross a governed boundary into a defined use context: local development, CI, staging, production runtime, infrastructure automation, model serving, regulated workflow, or developer endpoint.

\textbf{Custody envelope} means the set of controls that binds an artifact's identity, ingress route, execution authority, and revocation path strongly enough for the admitting institution to defend the admission decision.

\textbf{Use context} means the authority-bearing environment in which the artifact operates. The same artifact may be admissible in experimentation and inadmissible in production. The model scores artifact plus use context, not artifact type alone.

The relevant decision is not ``can a developer use this?'' The relevant decision is ``can the institution defend admitting this into infrastructure?''

\section{4. Related Work and Theoretical Lineage}\label{related-work-and-theoretical-lineage}

\subsection{4.1 Secure Consumption Frameworks}\label{secure-consumption-frameworks}

The nearest prior artifact is Microsoft's Secure Supply Chain Consumption Framework (S2C2F), now under the OpenSSF. S2C2F is consumption-focused rather than producer-focused. It addresses how organizations should securely consume open-source dependencies, defines solution-agnostic practices, organizes them into maturity levels, and begins with the problem of controlling artifact inputs. Its examples include source code, packages, modules, components, containers, libraries, and binaries, and its ``Ingest It'' practice is explicitly concerned with controlling the paths by which artifacts enter an organization.

The Custody Envelope Threshold is adjacent to S2C2F but does different work. S2C2F is a prescriptive maturity/practice framework: it tells organizations what secure consumption practices to implement. The Custody Envelope Threshold is a descriptive-predictive admission model: it explains why different artifact classes produce different institutional outcomes under comparable pressure. S2C2F asks whether an organization has implemented secure consumption practices. The Custody Envelope Threshold asks which custody dimension binds first, how that binding condition scales with execution authority, and which governance mode will emerge when direct admission cannot be defended.

The distinction matters. A package dependency, GitHub Action, Terraform provider, VS Code extension, and model artifact may all be ``open-source dependencies'' in a broad consumption sense. Institutionally, they delegate different forms of authority and therefore require different custody envelopes. S2C2F supplies a strong practice foundation. The Custody Envelope Threshold supplies the explanatory mechanism for divergence across artifact classes.

\subsection{4.2 Producer Integrity, Secure Development, and Attack Taxonomies}\label{producer-integrity-secure-development-and-attack-taxonomies}

SLSA and NIST SSDF are complementary but not equivalent to artifact admission. SLSA focuses on supply-chain integrity and the prevention of tampering in software artifacts and build processes. NIST SSDF organizes secure software-development practices that reduce vulnerability risk and provide a common vocabulary across producers and acquirers.

The academic supply-chain literature has similarly clarified the attack surface. Ohm et al.'s ``Backstabber's Knife Collection'' analyzed malicious packages in public ecosystems. Zahan et al.~studied weak links in npm package maintenance and maintainer account risk. Birsan's dependency-confusion work showed how public registry resolution can defeat private dependency assumptions. Ladisa et al.'s SoK taxonomized open-source software supply-chain attacks, incidents, and safeguards.

Those works primarily taxonomize attack vectors, weaknesses, or mitigations. The Custody Envelope Threshold taxonomizes institutional admission responses. It asks what institutions do with externally maintained artifacts after recognizing that supply-chain attacks are possible: admit, proxy, mediate, vendor-wrap, internalize, quarantine, or reject.

\subsection{4.3 Reference Monitor and Least Privilege}\label{reference-monitor-and-least-privilege}

The mechanism has an older security lineage. The reference-monitor concept holds that access requests should be validated against authorized relationships and that the validating mechanism should be tamper-resistant, always invoked, and small enough to analyze. Saltzer and Schroeder's design principles similarly emphasize complete mediation and least privilege: every access should be checked, and each program or user should operate with only the privileges required.

Artifact admission moves this logic upstream. Instead of asking only whether a runtime object may access a protected resource, the institution asks whether an externally maintained artifact may enter the system that will later build, deploy, operate, or mediate access to protected resources. A third-party CI action with repository secrets, a Terraform provider with cloud credentials, or a model artifact requiring remote code is not simply a file. It is an authority-bearing object. The admission decision should therefore be authority-adjusted.

\subsection{4.4 Transaction Cost Economics and Governance Modes}\label{transaction-cost-economics-and-governance-modes}

The outcome taxonomy also has a governance-economics lineage. Transaction cost economics treats governance as a comparative institutional choice under bounded rationality, opportunism, asset specificity, and transaction hazards. Institutions choose market, hybrid, or hierarchical governance modes depending on the hazards and coordination costs of the transaction.

Artifact admission is a boundary-of-the-firm problem translated into infrastructure. Direct admission resembles market contracting: the institution consumes an external artifact with minimal internal transformation. Proxying and policy mediation are hybrid modes: the artifact remains externally maintained, but the institution controls ingress, versioning, and execution policy. Vendor mediation transfers custody closure to a commercial or cloud intermediary. Internalization turns an external pattern into an internal artifact through copying, wrapping, republishing, or golden-path engineering. Quarantine or rejection occurs when transaction hazards cannot be governed at acceptable cost.

The Custody Envelope Threshold extends this lineage by adding delegated execution authority. Asset specificity explains why an organization may prefer internal or hybrid governance when substitution costs are high. Delegated authority explains why even low-substitution-cost artifacts may require strong governance when they touch secrets, infrastructure state, source code, production runtime, or model execution paths.

The six outcomes are therefore not arbitrary labels. They are governance modes for resolving custody deficits under delegated authority.

\section{5. The Instrument: Four Ordinal Conditions}\label{the-instrument-four-ordinal-conditions}

The framework becomes usable only when its conditions are measurable. The unit of analysis is an artifact in a use context. Each condition is scored on a 0-3 ordinal scale.

\subsection{5.1 Execution Authority}\label{execution-authority}

Execution authority is scored first because it determines the custody threshold the other dimensions must meet.

\begingroup\small
\begin{longtable}[]{@{}
  >{\raggedleft\arraybackslash}p{(\columnwidth - 6\tabcolsep) * \real{0.3077}}
  >{\raggedright\arraybackslash}p{(\columnwidth - 6\tabcolsep) * \real{0.2308}}
  >{\raggedright\arraybackslash}p{(\columnwidth - 6\tabcolsep) * \real{0.2308}}
  >{\raggedright\arraybackslash}p{(\columnwidth - 6\tabcolsep) * \real{0.2308}}@{}}
\toprule\noalign{}
\begin{minipage}[b]{\linewidth}\raggedleft
Score
\end{minipage} & \begin{minipage}[b]{\linewidth}\raggedright
Authority level
\end{minipage} & \begin{minipage}[b]{\linewidth}\raggedright
Operational definition
\end{minipage} & \begin{minipage}[b]{\linewidth}\raggedright
Examples
\end{minipage} \\
\midrule\noalign{}
\endhead
\bottomrule\noalign{}
\endlastfoot
0 & Non-executing / reference & Artifact is not executed, loaded, or interpreted in a way that can affect systems or data & Documentation, static examples, non-operational templates \\
1 & Local or low-impact execution & Artifact executes in a developer or test context without sensitive credentials, production effects, or privileged system access & Local dev utility, test-only dependency, sandboxed demo container \\
2 & Build, endpoint, or controlled runtime authority & Artifact runs in CI, developer endpoints, staging, or controlled runtime with access to source, build artifacts, limited secrets, or internal systems & CI action without deploy rights, editor extension, build plugin, staging image \\
3 & Privileged production or infrastructure authority & Artifact can affect production, cloud resources, deployment state, credentials, model-serving behavior, or regulated workflows & Terraform provider with cloud credentials, deployment action, production base image, Kubernetes operator, model requiring remote code in production \\
\end{longtable}
\endgroup

The execution authority score must be assigned to the use context, not the artifact label. A container image used in a disposable sandbox may score 1. The same image as a production base layer may score 3.

\subsection{5.2 Object Identity}\label{object-identity}

Object identity asks what exactly is being admitted.

\begingroup\small
\begin{longtable}[]{@{}
  >{\raggedleft\arraybackslash}p{(\columnwidth - 4\tabcolsep) * \real{0.4000}}
  >{\raggedright\arraybackslash}p{(\columnwidth - 4\tabcolsep) * \real{0.3000}}
  >{\raggedright\arraybackslash}p{(\columnwidth - 4\tabcolsep) * \real{0.3000}}@{}}
\toprule\noalign{}
\begin{minipage}[b]{\linewidth}\raggedleft
Score
\end{minipage} & \begin{minipage}[b]{\linewidth}\raggedright
Identity closure
\end{minipage} & \begin{minipage}[b]{\linewidth}\raggedright
Operational definition
\end{minipage} \\
\midrule\noalign{}
\endhead
\bottomrule\noalign{}
\endlastfoot
0 & Unstable or implicit identity & Artifact is referenced by an unpinned URL, mutable tag, branch, latest alias, public namespace alone, or other moving target \\
1 & Named but mutable identity & Artifact has a package name, action tag, model repo, module source, or image tag, but identity can change without institutional review \\
2 & Versioned or pinned identity & Artifact is bound to a version, digest, full commit SHA, checksum, model revision, or lockfile entry \\
3 & Verifiable and policy-bound identity & Artifact identity is pinned and supported by signatures, provenance, checksum verification, publisher trust, registry controls, or policy enforcement \\
\end{longtable}
\endgroup

Object identity is not naming. \texttt{ubuntu:latest}, \texttt{actions/foo@v1}, a Git branch, or a model repository without a pinned revision may name an artifact, but they do not sufficiently identify the artifact for high-authority admission.

\subsection{5.3 Ingress Path}\label{ingress-path}

Ingress path asks how the artifact enters the institution.

\begingroup\small
\begin{longtable}[]{@{}
  >{\raggedleft\arraybackslash}p{(\columnwidth - 4\tabcolsep) * \real{0.4000}}
  >{\raggedright\arraybackslash}p{(\columnwidth - 4\tabcolsep) * \real{0.3000}}
  >{\raggedright\arraybackslash}p{(\columnwidth - 4\tabcolsep) * \real{0.3000}}@{}}
\toprule\noalign{}
\begin{minipage}[b]{\linewidth}\raggedleft
Score
\end{minipage} & \begin{minipage}[b]{\linewidth}\raggedright
Ingress closure
\end{minipage} & \begin{minipage}[b]{\linewidth}\raggedright
Operational definition
\end{minipage} \\
\midrule\noalign{}
\endhead
\bottomrule\noalign{}
\endlastfoot
0 & Direct uncontrolled public fetch & Artifact is fetched directly from the public internet without institutional mediation \\
1 & Developer-controlled fetch with partial policy & Artifact enters through developer tooling with limited local checks or informal review \\
2 & Controlled institutional ingress & Artifact enters through an internal feed, proxy, remote repository, private registry, mirror, curated catalog, or approved marketplace \\
3 & Policy-enforced ingress boundary & Artifact can enter only through governed workflows with allowlists, scanning, approval, provenance checks, access controls, and audit logs \\
\end{longtable}
\endgroup

Ingress closure converts ``a developer pulled it from the internet'' into ``the institution admitted it through a governed path.''

\subsection{5.4 Revocation Capacity}\label{revocation-capacity}

Revocation capacity asks how quickly and completely the institution can block, remove, downgrade, replace, or prove non-use after admission changes.

\begingroup\small
\begin{longtable}[]{@{}
  >{\raggedleft\arraybackslash}p{(\columnwidth - 4\tabcolsep) * \real{0.4000}}
  >{\raggedright\arraybackslash}p{(\columnwidth - 4\tabcolsep) * \real{0.3000}}
  >{\raggedright\arraybackslash}p{(\columnwidth - 4\tabcolsep) * \real{0.3000}}@{}}
\toprule\noalign{}
\begin{minipage}[b]{\linewidth}\raggedleft
Score
\end{minipage} & \begin{minipage}[b]{\linewidth}\raggedright
Revocation closure
\end{minipage} & \begin{minipage}[b]{\linewidth}\raggedright
Operational definition
\end{minipage} \\
\midrule\noalign{}
\endhead
\bottomrule\noalign{}
\endlastfoot
0 & No inventory or block path & Institution cannot reliably identify where the artifact is used or prevent future use \\
1 & Manual repository-level revocation & Institution can remove or update the artifact manually in known locations but lacks central enforcement \\
2 & Central block or replacement path & Institution can block future fetches, update catalogs, rotate affected credentials, rebuild dependent artifacts, or enforce version replacement centrally \\
3 & Auditable rapid revocation & Institution can centrally identify usage, block execution or ingress, trigger rebuilds/redeployments, rotate exposed authority, and produce an audit record of revocation \\
\end{longtable}
\endgroup

Revocation is not deletion. It includes inventory, block path, credential response, downstream rebuild, cache invalidation, and proof that the old artifact no longer crosses the boundary.

\section{6. Closure Rule and Governance Selection}\label{closure-rule-and-governance-selection}

Let:

\begin{itemize}
\tightlist
\item
  \textbf{A} = execution authority score.
\item
  \textbf{I} = object identity score.
\item
  \textbf{G} = ingress path score.
\item
  \textbf{R} = revocation capacity score.
\end{itemize}

Direct custody closure is achieved when:

\[\min(I,G,R) \ge A\]

Equivalently: each custody condition must be at least as strong as the execution authority delegated to the artifact.

This is a weakest-link rule. The reason is structural. Object identity, ingress path, and revocation capacity do not substitute cleanly for one another. A signed artifact fetched through uncontrolled ingress can still bypass institutional admission. A proxied artifact referenced by a mutable tag can still change without review. A pinned artifact with no inventory or block path can still become unrecoverable after compromise. The envelope is only as strong as its weakest custody boundary.

The rule assumes commensurability across ordinal scores, but only at the level required for institutional decision-making. An identity score of 2 and a revocation score of 2 are not technically identical; they are comparable only as admission-strength levels. The instrument asks whether each dimension is weak, partially bounded, institutionally controlled, or policy-enforced relative to authority. It does not claim metric precision.

Authority and custody must remain orthogonal. \textbf{Authority} means what the artifact can do if executed, loaded, invoked, or deployed. \textbf{Custody} means how well the institution can bind what entered, through which route, and how it can be removed. A model artifact requiring \texttt{trust\_remote\_code=True} raises authority because code may execute; it also creates identity pressure because the remote code must be pinned and verified. Those are related observations, but the variables are distinct: authority concerns capability; custody concerns institutional control over the capability.

The A=0 boundary is deliberate. If an artifact has no execution or operational authority, the custody threshold is minimal. A static document, non-operational example, or inert reference artifact should not require the same admission process as a deployment action or Terraform provider.

Define the custody deficit:

\[D = A - \min(I,G,R)\], where D is treated as zero if the result is negative.

If $D=0$, direct admission is custody-closed. If $D>0$, direct admission is not defensible without additional controls or a different governance mode.

\subsection{6.1 Local Custody Specificity}\label{local-custody-specificity}

When direct custody closure fails, the next question is not simply which custody condition is deficient. It is whether the institution can close that deficit cheaply through generic controls, must rely on an external governance provider, or must internalize the artifact because safe use depends on local institutional knowledge.

This is the transaction cost economics term in the model. The relevant variable is \textbf{local custody specificity}: how much institution-specific knowledge is required to make the artifact safe to admit in the relevant use context.

\begingroup\small
\begin{longtable}[]{@{}
  >{\raggedleft\arraybackslash}p{(\columnwidth - 6\tabcolsep) * \real{0.3077}}
  >{\raggedright\arraybackslash}p{(\columnwidth - 6\tabcolsep) * \real{0.2308}}
  >{\raggedright\arraybackslash}p{(\columnwidth - 6\tabcolsep) * \real{0.2308}}
  >{\raggedright\arraybackslash}p{(\columnwidth - 6\tabcolsep) * \real{0.2308}}@{}}
\toprule\noalign{}
\begin{minipage}[b]{\linewidth}\raggedleft
Score
\end{minipage} & \begin{minipage}[b]{\linewidth}\raggedright
Specificity level
\end{minipage} & \begin{minipage}[b]{\linewidth}\raggedright
Operational definition
\end{minipage} & \begin{minipage}[b]{\linewidth}\raggedright
Typical governance implication
\end{minipage} \\
\midrule\noalign{}
\endhead
\bottomrule\noalign{}
\endlastfoot
0 & Commodity & Safe use depends mostly on generic ecosystem controls; the artifact does not need to know the institution's architecture, policy, or operating model & Proxying, cataloging, or vendor mediation can scale efficiently \\
1 & Workflow-specific & Safe use depends on a standard workflow or toolchain, but ordinary policy controls can govern it & Policy mediation or curated catalogs are usually sufficient \\
2 & Platform-specific & Safe use depends on the institution's platform architecture, identity model, runtime model, deployment pattern, or credential boundary & Vendor mediation, platform wrapping, or internal templates may be needed \\
3 & Institution-specific & Safe use requires local institutional knowledge that cannot be supplied reliably by the upstream artifact alone & Internalization is often the defensible mode \\
\end{longtable}
\endgroup

This anchor is deliberately intrinsic. It does not ask whether the artifact was internalized. It asks how much local institutional knowledge is required for safe use before the outcome is coded.

This distinction separates Terraform providers from Terraform modules. A provider may have high execution authority but lower local custody specificity: it speaks to a cloud API and can be governed through provider signing, lockfiles, mirrors, credentials, and workspace policy. A public module may have lower direct execution authority but higher specificity: safe use depends on local network assumptions, IAM conventions, encryption defaults, tagging standards, observability requirements, and cost or compliance posture. The provider can often be admitted through distribution custody. The module often has to be wrapped or internalized.

\subsection{6.2 Governance-Selection Heuristic}\label{governance-selection-heuristic}

A formal optimization claim would overstate what can be measured from public evidence. Cross-sectional public documentation rarely lets an analyst prove that an observed governance mode is the true cost-minimizing mode.

The paper therefore uses a \textbf{comparative governance heuristic}, not a cost computation.

The selection rule is:

\textbf{When direct custody closure fails, the institution selects the lowest-burden governance mode that appears capable of closing the binding deficit at an acceptable risk/value tradeoff, given the artifact's authority and local custody specificity.}

The model uses three observable inputs:

\begin{enumerate}
\def\labelenumi{\arabic{enumi}.}
\tightlist
\item
  \textbf{Binding deficit:} which custody dimension falls below authority first.
\item
  \textbf{Local custody specificity:} how much institution-specific knowledge is needed for safe use.
\item
  \textbf{External closure availability:} whether an external platform, marketplace, registry, vendor, or cloud provider can plausibly supply the missing custody boundary.
\end{enumerate}

The governance prediction is heuristic but falsifiable: the predicted primary mode must be stated before the observed mode is coded.

\subsection{6.3 Derived Governance Modes}\label{derived-governance-modes}

\begingroup\small
\begin{longtable}[]{@{}
  >{\raggedright\arraybackslash}p{(\columnwidth - 4\tabcolsep) * \real{0.3333}}
  >{\raggedright\arraybackslash}p{(\columnwidth - 4\tabcolsep) * \real{0.3333}}
  >{\raggedright\arraybackslash}p{(\columnwidth - 4\tabcolsep) * \real{0.3333}}@{}}
\toprule\noalign{}
\begin{minipage}[b]{\linewidth}\raggedright
Binding condition
\end{minipage} & \begin{minipage}[b]{\linewidth}\raggedright
Specificity / external-closure profile
\end{minipage} & \begin{minipage}[b]{\linewidth}\raggedright
Predicted primary governance mode
\end{minipage} \\
\midrule\noalign{}
\endhead
\bottomrule\noalign{}
\endlastfoot
No deficit: $\min(I,G,R) \ge A$ & Closure already sufficient & Admitted \\
Ingress below authority; low specificity; high usage pressure & Internal feed, proxy, remote repository, or mirror can close ingress cheaply & Proxied \\
Identity below authority; policy controls can close the gap & Pinning, full-SHA references, signatures, lockfiles, allowlists, publisher trust, permission limits & Policy-mediated \\
Multiple deficits; low-to-medium specificity; credible external closure provider exists & Cloud platform, managed catalog, marketplace, gateway, audit layer, deployment wrapper, support/contract layer & Vendor-mediated \\
Deficit depends on local platform or institutional knowledge & Copy, wrap, republish, fork, convert into golden path, private module, internal tool, or internal base line & Internalized \\
Deficit remains high and available closure modes exceed value/risk tolerance & Experiment only, sandbox only, production block, or ban & Quarantined/rejected \\
\end{longtable}
\endgroup

This formulation keeps the cost/specificity insight without overclaiming. The closure rule determines whether direct admission fails. The governance heuristic predicts how the institution responds.

\subsection{6.4 Co-Binding and Primary-Mode Tie-Breaks}\label{co-binding-and-primary-mode-tie-breaks}

The weakest-link rule identifies whether direct custody closure fails:

\[\min(I,G,R) < A\]

In some cases, more than one custody dimension co-binds. A public production base image may have weak ingress and weak revocation at the same time. A raw high-authority MCP server may have weak identity, ingress, and revocation simultaneously. A third-party GitHub Action referenced by a mutable tag may also show a three-way deficit: weak identity, weak ingress, and weak revocation.

The coding protocol therefore needs a deterministic tie-break rule for assigning a single primary predicted governance mode.

Let \textbf{B} be the set of custody dimensions tied at the minimum score and below authority:

\[B = \{I,G,R \mid \text{score}=\min(I,G,R) \text{ and } \text{score}<A\}\]

If \textbf{B} contains one dimension, the primary mode follows the ordinary mapping:

\begin{itemize}
\tightlist
\item
  identity deficit -\textgreater{} policy-mediated;
\item
  ingress deficit -\textgreater{} proxied;
\item
  revocation deficit -\textgreater{} curated, vendor-mediated, internalized, or rejected, depending on specificity and available closure.
\end{itemize}

If \textbf{B} contains multiple dimensions, the primary mode follows the deficit whose closure cascades to the others. A \textbf{cascade closure} is a control move that closes one binding deficit and materially enables or closes another. The primary predicted mode is assigned to the governance layer that supplies the first cascade closure.

\begingroup\small
\begin{longtable}[]{@{}
  >{\raggedright\arraybackslash}p{(\columnwidth - 4\tabcolsep) * \real{0.3333}}
  >{\raggedright\arraybackslash}p{(\columnwidth - 4\tabcolsep) * \real{0.3333}}
  >{\raggedright\arraybackslash}p{(\columnwidth - 4\tabcolsep) * \real{0.3333}}@{}}
\toprule\noalign{}
\begin{minipage}[b]{\linewidth}\raggedright
Co-binding pattern
\end{minipage} & \begin{minipage}[b]{\linewidth}\raggedright
Cascade question
\end{minipage} & \begin{minipage}[b]{\linewidth}\raggedright
Primary-mode rule
\end{minipage} \\
\midrule\noalign{}
\endhead
\bottomrule\noalign{}
\endlastfoot
Identity + revocation & Does identity closure create usable inventory and a block path? & If pinning, signing, lockfiles, full-SHA references, publisher trust, or allowlists create the inventory needed for revocation, predict policy-mediated. If revocation requires local rebuild or ownership, predict internalized. \\
Ingress + revocation & Does closing ingress require changing the artifact's route, or can the existing route be retained and constrained by policy? & If closing ingress requires a new governed route - internal feed, registry, mirror, proxy, private marketplace, curated catalog, or route-level chokepoint - predict proxied. If the artifact can continue entering through the existing platform route but ingress is closed by allowlists, publisher constraints, version constraints, permission policy, or endpoint policy, predict policy-mediated. \\
Identity + ingress & Can identity be closed inside the existing platform control plane? & If pinning, signing, lockfiles, full-SHA references, publisher trust, or allowlists close identity without changing the distribution route, predict policy-mediated. If direct public ingress remains unacceptable even after identity closure, predict proxied. \\
Identity + ingress + revocation & Is there an existing control plane whose policy controls can close all three? & If pinning, signing, allowlisting, permission scoping, lockfiles, publisher trust, or platform policy can close the deficits inside the existing control plane, predict policy-mediated. If the first necessary move is a route/catalog/proxy boundary, predict proxied. If multiple deficits require an external platform wrapper and $S \le 2$, predict vendor-mediated. If safe use requires institution-specific knowledge and S = 3, predict internalized. If no available governance mode can close the deficit at acceptable risk/value, predict quarantined/rejected. \\
Any co-binding case with S = 3 & Does safe use require local institutional knowledge? & Predict internalized, unless the artifact is limited to experimentation or a trusted internal platform already supplies equivalent custody. \\
\end{longtable}
\endgroup

This rule prevents post hoc mode selection. The model must emit one primary predicted governance mode before the observed primary mode is coded. Secondary controls may be reported, but they cannot rescue a failed primary prediction.

The governing principle is:

\textbf{In a multi-deficit case, the primary governance mode follows the deficit whose closure cascades to the others.}

For arbitrary public production base images, ingress and revocation co-bind. Closing ingress through a registry, mirror, proxy, or curated catalog prevents uncontrolled future pulls and creates the inventory surface needed for revocation. The primary mode is therefore proxied; curation, digest pinning, rebuilds, and runtime admission are secondary controls.

For third-party GitHub Actions, identity, ingress, and revocation can co-bind. But full-SHA pinning, allowlisting, least-privilege permissions, and organization policy can often close the relevant deficits inside the existing GitHub control plane. Identity closure cascades into revocation because pinned references create inventory and blockability. The primary mode is therefore policy-mediated, not vendor-mediated.

For VS Code extensions on managed endpoints, ingress and revocation may co-bind, but if the primary observed control is an allowlist, extension policy, publisher constraint, or version constraint applied inside the existing endpoint-management or marketplace-control plane, the mode is policy-mediated. If the artifact may enter only through a private marketplace or internal extension catalog, the mode is proxied.

\subsection{6.5 Catalog / Marketplace Convention}\label{catalog-marketplace-convention}

A curated internal catalog or private marketplace can function both as an ingress route and as a policy gate. To keep coding consistent:

\begin{itemize}
\tightlist
\item
  \textbf{Catalog-as-route -\textgreater{} proxied.} If the artifact is allowed only because it enters through a private marketplace, internal feed, registry, mirror, proxy, or curated catalog, code the primary mode as proxied.
\item
  \textbf{Constraint-on-otherwise-admitted artifact -\textgreater{} policy-mediated.} If the artifact can enter through the ordinary platform route but only when it satisfies pinning, signing, allowlisting, publisher, permission, or version constraints, code the primary mode as policy-mediated.
\item
  \textbf{External platform wrapper -\textgreater{} vendor-mediated.} If the artifact is acceptable only because a cloud provider, commercial platform, gateway, managed catalog, or vendor service supplies custody boundaries the institution does not itself supply, code the primary mode as vendor-mediated.
\item
  \textbf{Local ownership required -\textgreater{} internalized.} If safe use requires copying, wrapping, republishing, or converting the artifact into an institution-owned artifact, code the primary mode as internalized.
\end{itemize}

This convention should be applied before coding the observed outcome.

\subsection{6.6 Scrutiny as the Positive-Claim Moderator}\label{scrutiny-as-the-positive-claim-moderator}

The model's normative claim is not enough to predict institutional behavior. Many organizations operate with weak custody because no effective scrutiny has yet forced the issue. The positive claim requires an observable moderator: \textbf{institutional scrutiny}.

Scrutiny means exposure to recurring external or internal review that makes weak-custody, high-authority artifact admission difficult to defend. It should be coded before observing the admission outcome.

\begingroup\small
\begin{longtable}[]{@{}
  >{\raggedleft\arraybackslash}p{(\columnwidth - 4\tabcolsep) * \real{0.4000}}
  >{\raggedright\arraybackslash}p{(\columnwidth - 4\tabcolsep) * \real{0.3000}}
  >{\raggedright\arraybackslash}p{(\columnwidth - 4\tabcolsep) * \real{0.3000}}@{}}
\toprule\noalign{}
\begin{minipage}[b]{\linewidth}\raggedleft
Score
\end{minipage} & \begin{minipage}[b]{\linewidth}\raggedright
Scrutiny level
\end{minipage} & \begin{minipage}[b]{\linewidth}\raggedright
Observable proxy
\end{minipage} \\
\midrule\noalign{}
\endhead
\bottomrule\noalign{}
\endlastfoot
0 & Low scrutiny & Individual, hobby, early-stage, unmanaged, or informal development context; no visible external assurance obligations \\
1 & Moderate operational scrutiny & Organization has production users, internal security review, basic customer assurance, or formal platform ownership \\
2 & High enterprise scrutiny & SOC 2, ISO 27001, customer security questionnaires, enterprise procurement review, cyber-insurance review, regulated customers, or formal third-party risk process \\
3 & Regulated / mission-critical scrutiny & FedRAMP, PCI, HIPAA, financial-services supervision, critical infrastructure, government workloads, export-controlled workflows, or public post-incident remediation obligations \\
\end{longtable}
\endgroup

The positive claim is now:

\textbf{For the same artifact class and authority level, higher-scrutiny institutions should exhibit stronger custody closure or more restrictive governance modes.}

This prevents scrutiny from becoming an unfalsifiable escape clause. A case cannot be dismissed after the fact as ``not serious.'' The scrutiny score must be assigned ex ante from observable institutional features.

\subsection{6.7 External Closure Availability}\label{external-closure-availability}

Governance predictions are conditional on the contemporaneous external-closure market.

A governance mode becomes available only when some actor can supply the missing custody boundary. A model artifact may be quarantined when no credible model catalog, scanner, managed endpoint, or support layer exists; the same artifact may become vendor-mediated once a cloud provider or enterprise hub supplies identity, scanning, logging, isolation, and revocation. A high-authority MCP server may be rejected when the only available form is raw local installation; the same server may become vendor-mediated once gateways, portals, scoped authorization, tool-call logs, and DLP controls become available.

This is not an exception to the model. It is part of the governance selector. The predicted primary mode is conditional on the available closure market at the time of observation. That is why temporal validity matters: as external providers add new custody surfaces, artifact classes can move from quarantine to vendor mediation, or from vendor mediation to direct admission if ecosystem-level identity, ingress, and revocation controls mature.

\section{7. Case Selection and Analytic Method}\label{case-selection-and-analytic-method}

The derivation set consists of six artifact classes chosen because they face the same structural pressure but exhibit divergent institutional outcomes:

\begin{enumerate}
\def\labelenumi{\arabic{enumi}.}
\tightlist
\item
  package-registry dependencies;
\item
  GitHub Actions;
\item
  container images and base images;
\item
  Terraform providers and modules;
\item
  developer-environment extensions;
\item
  open model artifacts.
\end{enumerate}

The cases satisfy six criteria. Each involves externally maintained artifacts entering institutional infrastructure. Each is developer-accessible through low-friction tooling. Each can cross from experimentation into authority-bearing environments. Each has public controls, documentation, or incidents that reveal admissibility pressure. Each differs from the others in which custody dimension binds first. Each matters to procurement-adjacent infrastructure decision-makers.

The analysis below is an instrumented comparative case analysis, not a completed archival validation study. It demonstrates that the rubric can classify real artifact classes and generate differentiated predictions. A full empirical validation would require a larger corpus of public admission artifacts - organization policies, registry configurations, CI allowlists, private marketplace rules, model-catalog eligibility requirements, and incident-response playbooks - coded independently against the rubric with reported inter-rater reliability. That validation design is specified later.

\section{8. Comparative Application}\label{comparative-application}

\subsection{8.1 Package-Registry Dependencies: Proxied Admission}\label{package-registry-dependencies-proxied-admission}

Package-registry dependencies are the clearest case where usefulness and direct admissibility diverge. Modern software depends on public registries such as npm, PyPI, Maven Central, RubyGems, and NuGet. Banning them wholesale is usually unrealistic. But direct public resolution is difficult to defend because package names, maintainer accounts, version publication, dependency metadata, and transitive updates are controlled outside the institution.

The dependency-confusion pattern makes the admission problem explicit. An organization may rely on private packages from an internal registry while an attacker registers the same package name on a public registry; default resolution behavior may fetch the public package unless scopes, registry configuration, or other controls prevent it.

Scored against the instrument, a typical public package used in production has moderate authority. It may not deploy infrastructure by itself, but it can execute in build and runtime paths. Its public identity may be versioned, but dependency graphs are broad, namespaces are externally controlled, and transitive updates create revocation problems. Direct public fetch often leaves ingress below the authority threshold.

The institutional response is proxied admission. Public packages enter through internal feeds, artifact proxies, remote repositories, internal mirrors, curated catalogs, lockfiles, vulnerability gates, registry signatures, and provenance mechanisms. Google Artifact Registry remote repositories act as proxies for upstream sources and can cache packages while combining with virtual repositories to mitigate dependency-confusion risk. Azure Artifacts centralizes packages in feeds and gives feed owners control over externally sourced public versions. npm registry signatures and trusted publishing strengthen object identity and publication provenance.

The outcome is not ``packages are trusted.'' The outcome is that package dependencies are admitted through resolver governance. Their public form is not the institutional artifact. The institutional artifact is the version that passed through the organization's ingress path, identity controls, vulnerability process, and revocation mechanisms.

\subsection{8.2 GitHub Actions: Policy-Mediated or Quarantined}\label{github-actions-policy-mediated-or-quarantined}

GitHub Actions expose the execution-authority condition more sharply than language dependencies.

At the developer surface, a marketplace action looks like a reusable dependency. Institutionally, it is delegated CI code execution. It may run inside workflows that have access to repository contents, generated artifacts, package-publishing credentials, cloud tokens, deployment secrets, and internal network paths.

GitHub's own guidance treats this as an authority problem. Pinning an action to a full-length commit SHA is described as the only way to use an action as an immutable release; tags can move, and compromised actions can access secrets and tokens available to the workflow. GitHub also provides policies that can require full-SHA pinning. The \texttt{tj-actions/changed-files} compromise illustrates the mechanism: mutable tags were retroactively changed to point to a malicious commit, and CI/CD secrets were exposed in workflow logs across a large population of affected repositories.

A third-party action referenced by a mutable tag may have A=2 or A=3 depending on secrets and deployment permissions, I=1, G=1, and R=1. That fails the closure rule. The same action pinned to a full commit SHA, allowed by organization policy, reviewed, constrained by least-privilege \texttt{GITHUB\_TOKEN} permissions, and monitored for revocation may move to I=2 or 3, G=2 or 3, and R=2.

The outcome is policy-mediated admission. Some actions are allowed: first-party actions, internal organization actions, reviewed third-party actions, pinned actions, or actions operating under least-privilege permissions. Mutable, unknown, or high-authority actions remain quarantined or blocked.

The framework explains why GitHub Actions can be treated more strictly than package dependencies inside the same institution. Package dependencies can often be resolved through internal registries and rebuilt artifacts. A CI action can influence the process that builds, signs, publishes, deploys, or releases software. Its authority is not merely application-level. It is production-process authority.

\subsection{8.3 Container Images: Proxied Admission with Curated Runtime Controls}\label{container-images-proxied-admission-with-curated-runtime-controls}

Container images demonstrate the interaction between ingress, authority, and revocation capacity.

A container image is easy to pull and hard to defend if its source, base lineage, update cadence, vulnerabilities, and runtime behavior are unknown. A base image is especially consequential because it becomes an inherited substrate. It is copied into downstream images, embedded in CI builds, cached in registries, deployed across services, and sometimes forgotten until a vulnerability or end-of-life notice forces rebuilds.

Docker itself distinguishes trust surfaces. Docker Official Images are curated and maintained according to program criteria. Docker Trusted Content distinguishes Official Images, Hardened Images, Verified Publisher images, and Docker-Sponsored Open Source Software images. That distinction matters because Docker Hub is not a single trust domain.

Scored against the instrument, an arbitrary public image used in production may have A=3, I=1 or 2 depending on digest pinning, G=0 or 1 if pulled directly, and R=0 or 1 if there is no central rebuild or block path. That fails. A curated base image pulled through an internal registry, pinned by digest, scanned, signed, admitted through deployment policy, and centrally rebuildable can satisfy the threshold. Sigstore-style admission policies that validate signatures and attestations at cluster boundaries further illustrate how runtime admission becomes custody enforcement.

This produces \textbf{proxied admission with curated runtime controls}. For arbitrary public production base images, ingress and revocation commonly co-bind: the institution must first prevent uncontrolled future pulls through a registry, mirror, or proxy, then add curation, digest pinning, scanning, rebuild procedures, and runtime admission policy. Under the tie-break rule, proxying is the primary governance mode; curated base lines and runtime policy are secondary controls that complete the envelope.

The crucial distinction is between a container image as a developer convenience and a container image as production substrate. The same artifact type crosses the threshold in one context and fails it in another.

\subsection{8.4 Terraform Providers and Modules: Same Toolchain, Different Admission Mechanics}\label{terraform-providers-and-modules-same-toolchain-different-admission-mechanics}

Terraform provides a useful internal comparison because providers and modules appear adjacent but bind on different conditions.

A Terraform provider is high-authority. It is a plugin that interacts with upstream APIs, often cloud APIs with permissions to create, modify, or destroy infrastructure. On execution authority alone, providers look dangerous.

Yet Terraform providers have unusually strong custody mechanics. Providers installed from the Terraform Registry are cryptographically signed and verified during installation. Terraform's dependency lock file records provider selections and checksums and is intended to be committed for review. Terraform supports provider mirrors, including local filesystem and network mirrors, allowing controlled environments to install providers without direct upstream registry access.

Those controls make provider admission governable. The institution can ask which provider address, which version, which signature, which checksum, which mirror, which workspace, which credentials, and which review path. A high-authority provider can become admissible because identity, ingress, and revocation capacity can be raised toward the authority threshold.

Terraform modules are different. Modules can be sourced from registries, Git repositories, object storage, or local files. Git-based sources may be pinned to a tag, branch, or SHA. But Terraform's dependency lock file tracks providers, not remote module version selections. Exact module version constraints and registry practices must carry the repeatability burden for modules.

The deeper distinction is institutional. Providers are high-authority binaries with mature distribution custody. Modules are architectural artifacts. They encode network shape, IAM assumptions, tagging conventions, encryption defaults, observability patterns, and cost-bearing resource choices. A module may be technically harmless as code while institutionally inadmissible as architecture.

The outcome is split. Providers can be admitted through signed, locked, mirrored distribution. Public modules more often become internalized: copied into private module registries, wrapped by platform teams, reviewed against internal standards, or converted into golden modules. The same toolchain produces different admission mechanics because different custody conditions bind first.

\subsection{8.5 Developer Extensions: Endpoint Artifact Custody}\label{developer-extensions-endpoint-artifact-custody}

Developer-environment extensions sit in a blind spot between productivity tooling and infrastructure trust.

A VS Code extension is not usually perceived as production infrastructure. It is installed by a developer, runs in an editor, and improves workflow. But the editor is where source code, terminals, local credentials, language servers, AI agents, cloud CLIs, package managers, internal repositories, and external services meet. An extension can therefore sit near sensitive material even if it never runs in production.

Microsoft's Workspace Trust documentation makes the boundary visible: Restricted Mode disables or limits features such as terminal, tasks, debugging, workspace settings, extensions, and AI agents, but Workspace Trust cannot prevent a malicious extension from executing code and ignoring Restricted Mode. Enterprise extension controls allow administrators to define allowed extensions by publisher, extension ID, version, and platform, and private extension marketplaces can self-host, upstream, or rehost extensions under enterprise-specific controls. Marketplace-level trust measures such as publisher signing and security-sensitive behavior detection improve identity and authenticity, but they do not settle admission by themselves.

Scored against the instrument, a developer extension with access to source code, terminal execution, external requests, AI-agent behavior, or local credentials may reach A=2. If installed directly from a public marketplace with automatic updates and no central inventory, identity, ingress, and revocation may sit below that threshold. Enterprise allowlists and private marketplaces raise ingress and revocation capacity. Publisher signing and version constraints raise object identity.

The outcome is developer-surface catalog admission when the route changes, or policy-mediated admission when the marketplace route is retained but constrained by allowlists, publisher constraints, version constraints, or endpoint policy. Individual developers experience the policy as an extension allowlist. The institution experiences it as endpoint artifact custody.

The case matters because AI-assisted development raises endpoint authority. Extensions that invoke tools, operate terminals, call models, modify files, or connect to external systems are no longer cosmetic. They become part of the software production environment.

\subsection{8.6 Open Model Artifacts: Vendor-Mediated or Quarantined}\label{open-model-artifacts-vendor-mediated-or-quarantined}

Open model artifacts expose the boundary of conventional software-supply-chain controls.

A model repository can contain weights, configuration, tokenizer files, custom modeling code, evaluation material, datasets, and deployment instructions. Some files are inert data. Some are executable or deserialized through dangerous formats. Some require remote code. Some carry licensing constraints. Some have uncertain training provenance. Some are safe to inspect and unsafe to serve.

Hugging Face runs malware scanning on repository files and separately documents pickle scanning, including ClamAV scans and pickle import scans. Its Transformers documentation warns that models requiring \texttt{trust\_remote\_code=True} require verification of modeling files and recommends pinning a revision to protect against repository updates. Microsoft Foundry and Azure Machine Learning provide model-catalog and deployment paths; Hugging Face's Azure integration imposes eligibility requirements, restrictions on \texttt{trust\_remote\_code}, Safetensors requirements, and mandatory security scanning before model import into customer tenancy.

Scored against the instrument, a raw model artifact used in experimentation may have A=1 or 2. The same artifact served in production, loaded with remote code, or integrated into regulated workflows may have A=3. Public hub identity may be insufficient without pinned revision and file-level controls. Direct public fetch may fail ingress closure. Revocation may be difficult if models are cached, embedded into pipelines, deployed to endpoints, or fine-tuned into derivatives.

The outcome is increasingly model-catalog mediation. Raw hubs remain discovery surfaces. Production admission flows toward catalog identity, scanning, deployment isolation, managed endpoints, support expectations, access logs, and revocation handles.

This does not mean institutions generically trust Microsoft, Hugging Face, or any cloud provider. It means raw model artifacts often lack sufficient custody closure for high-authority production paths, so a platform layer supplies the missing envelope.

\section{9. Binding Deficits Across the Derivation Cases}\label{binding-deficits-across-the-derivation-cases}

The comparative pattern can be summarized as follows.

\begingroup\small
\begin{longtable}[]{@{}
  >{\raggedright\arraybackslash}p{(\columnwidth - 6\tabcolsep) * \real{0.2500}}
  >{\raggedright\arraybackslash}p{(\columnwidth - 6\tabcolsep) * \real{0.2500}}
  >{\raggedright\arraybackslash}p{(\columnwidth - 6\tabcolsep) * \real{0.2500}}
  >{\raggedright\arraybackslash}p{(\columnwidth - 6\tabcolsep) * \real{0.2500}}@{}}
\toprule\noalign{}
\begin{minipage}[b]{\linewidth}\raggedright
Artifact class
\end{minipage} & \begin{minipage}[b]{\linewidth}\raggedright
Typical high-value use context
\end{minipage} & \begin{minipage}[b]{\linewidth}\raggedright
Binding deficit
\end{minipage} & \begin{minipage}[b]{\linewidth}\raggedright
Predicted governance mode
\end{minipage} \\
\midrule\noalign{}
\endhead
\bottomrule\noalign{}
\endlastfoot
Package dependencies & Build/runtime dependency graphs & Ingress and namespace resolution & Proxied admission \\
GitHub Actions & CI/CD with tokens, secrets, or release authority & Mutable identity plus high authority & Policy-mediated or quarantined \\
Container images & Production runtime and base-image lineage & Revocation and ingress & Proxied admission with curated runtime controls \\
Terraform providers & Infrastructure automation with cloud credentials & High authority, mitigated by strong identity and mirrors & Admitted with signed/locked/mirrored distribution \\
Terraform modules & Infrastructure design patterns & Local custody specificity and weak module repeatability & Internalized or private-registry mediated \\
Developer extensions & Source-code and endpoint environment & Endpoint authority and update control & Policy-mediated or catalog-mediated \\
Model artifacts & Production model serving or remote-code loading & Unsafe loading, catalog identity, serving revocation & Vendor-mediated or quarantined \\
\end{longtable}
\endgroup

The framework's explanatory compression is that these outcomes are not separate stories. They are different failures or satisfactions of the same threshold rule. Different conditions bind first.

\section{10. Stress Tests and Boundary Conditions}\label{stress-tests-and-boundary-conditions}

\subsection{\texorpdfstring{10.1 The \texttt{curl\ \textbar{}\ bash} Counterexample}{10.1 The curl \textbar{} bash Counterexample}}\label{the-curl-bash-counterexample}

The sharpest counterexample is the \texttt{curl\ \textbar{}\ bash} installation pattern. It is ubiquitous in developer documentation, often executes with high local authority, has weak object identity, uncontrolled ingress, and poor revocation. The model appears to predict quarantine or rejection, yet the practice persists.

The counterexample is useful because it separates developer practice from institutional admission. In unmanaged or low-maturity contexts, \texttt{curl\ \textbar{}\ bash} persists because no institutional admission boundary is being enforced. In mature contexts, the pattern is commonly transformed: scripts are inspected, pinned, mirrored, repackaged, executed in disposable environments, converted into internal installers, or blocked in production build paths. S2C2F's examples of artifact ingestion include \texttt{curl}, direct public package-manager use, copy-pasted source, and binaries in repositories precisely because uncontrolled ingress is a consumption problem requiring standard governance.

The model therefore treats \texttt{curl\ \textbar{}\ bash} as a stress-confirming case if it appears in low-authority experimentation or immature governance, and as a falsifying case if high-scrutiny institutions normalize it in high-authority production paths without mediation.

\subsection{10.2 Artifacts Rejected Despite Closed Custody}\label{artifacts-rejected-despite-closed-custody}

Some artifacts may be rejected despite satisfying identity, ingress, authority, and revocation closure. A model may be disallowed because of licensing, training-data provenance, export control, privacy, content-safety, or sector-specific regulatory concerns. A cryptographically verified artifact may still be rejected because its license is unacceptable. A dataset may be rejected because its consent basis is insufficient.

These are not custody failures. They are separate downstream gates. The framework does not absorb them into a growable custody condition. Doing so would make the model unfalsifiable. The scope is technical-institutional admission of an external artifact into governed infrastructure. Legal, semantic, behavioral, export, and business-purpose adequacy may still block use after custody closure. They should be modeled separately.

\subsection{10.3 Artifacts Admitted Despite Weak Closure}\label{artifacts-admitted-despite-weak-closure}

An artifact apparently admitted despite weak closure may not actually be an anomaly if its execution authority is low. A public package used in a disposable test harness, a model pulled for offline research, or an extension installed in an isolated sandbox can have weak custody without violating the threshold because the authority score is low. The model becomes strained only when authority is high and custody remains weak.

This is why use context is the unit of analysis. Artifact type alone is insufficient.

\subsection{10.4 Outcome Overlap}\label{outcome-overlap}

The outcome taxonomy is intentionally composable. A package can be proxied and policy-mediated. A model can be vendor-mediated and quarantined outside that path. A container image can be curated and policy-enforced at runtime.

For prediction, the operative question is the binding constraint: which control mode is necessary before admission becomes defensible? The binding-constraint framing avoids treating governance modes as mutually exclusive boxes.

\section{11. Validation Design}\label{validation-design}

The framework can be tested, but only if the unit of analysis and dependent variable are separated correctly.

The unit of analysis is:

\textbf{raw artifact class x use context x institutional control surface}

The unit is not ``npm through an internal feed'' or ``model through a managed catalog,'' because those phrases already contain the outcome. The correct unit is an artifact in a use context, such as:

\begin{itemize}
\tightlist
\item
  npm package dependency in production build/runtime;
\item
  third-party GitHub Action in CI with repository secrets;
\item
  public container image used as a production base image;
\item
  Terraform provider used in cloud infrastructure automation;
\item
  public Terraform module used for production network provisioning;
\item
  VS Code extension on a managed developer endpoint;
\item
  Hugging Face model artifact requiring remote code in production serving;
\item
  MCP server with write access to an enterprise system.
\end{itemize}

For each unit, coders score A, I, G, R, and S from the controls present in the institutional setting. Separately, coders assign a primary observed governance mode using the dependent-variable rubric. The model then predicts the primary governance mode from the custody deficit and the cost/specificity selector.

The four preregistered hypotheses are:

\begin{itemize}
\tightlist
\item
  \textbf{H1 (Authority scaling).} Higher execution-authority artifacts are associated with stronger custody controls, where custody strength is $\min(I,G,R)$.
\item
  \textbf{H2 (Binding-deficit prediction).} The deterministic Custody Envelope mapping predicts the primary observed governance mode better than a majority-class baseline.
\item
  \textbf{H3 (Context dependence).} The same artifact class receives different predicted and observed governance modes when used in different authority contexts.
\item
  \textbf{H4 (Scrutiny interaction).} The effect of scrutiny on custody strength is larger at higher execution authority; in cross-section the interaction term $A \times Q$ should be positive when predicting custody strength or governance-mode restrictiveness.
\end{itemize}

\subsection{11.1 Pre-Governance and Realized Custody Measures}\label{pre-governance-and-realized-custody-measures}

The empirical pilot distinguishes pre-governance custody affordances from realized custody controls.

\textbf{Pre-governance scores} are used as inputs to the deterministic prediction function. They describe the artifact's default-path custody posture and platform affordances before the institution's realized admission choice is coded.

Pre-governance variables include:

\begin{itemize}
\tightlist
\item
  A - Execution Authority;
\item
  I\_pre - Object Identity in the artifact's default path;
\item
  G\_pre - Ingress Path in the artifact's default path;
\item
  R\_pre - Revocation Capacity in the artifact's default path;
\item
  S - Local Custody Specificity;
\item
  Q - Institutional Scrutiny;
\item
  Route-change vs in-place-policy affordance;
\item
  Control-plane closability;
\item
  External-closure availability.
\end{itemize}

The deterministic prediction function uses A, I\_pre, G\_pre, R\_pre, S, Q, and affordance variables to emit one primary predicted governance mode.

\textbf{Realized custody scores} are used for H1 and H4. They describe the custody controls the institution actually applies at the admission point.

Realized variables include:

\begin{itemize}
\tightlist
\item
  I\_realized - Object Identity under the institution's applied controls;
\item
  G\_realized - Ingress Path under the institution's applied controls;
\item
  R\_realized - Revocation Capacity under the institution's applied controls;
\item
  $\mathrm{CustodyStrength}_{\mathrm{realized}} = \min(I_{\mathrm{realized}},G_{\mathrm{realized}},R_{\mathrm{realized}})$;
\item
  Primary observed governance mode;
\item
  Governance-mode restrictiveness.
\end{itemize}

H2 tests whether pre-governance custody deficits predict the primary observed governance mode. H1 and H4 use realized custody strength and/or governance-mode restrictiveness as dependent variables. They do not use pre-governance $\min(I_{\mathrm{pre}},G_{\mathrm{pre}},R_{\mathrm{pre}})$ as the primary custody-strength outcome.

Coders must not infer pre-governance scores from the institution's observed governance choice. Pre-governance scores should be based on the artifact's default path and technical affordances. Realized scores should be based on what the institution actually permits, requires, blocks, or mediates.

\subsection{11.2 Point Scores, Not Ranges}\label{point-scores-not-ranges}

The validation study uses point scores, not score ranges.

Illustrative tables may show ranges because public artifact classes vary across institutions and use contexts. Those ranges are explanatory, not predictive. They show where the same artifact class can move under different authority, scrutiny, or custody conditions.

In the coded study, each observation must receive a single point score for A, I, G, R, S, and Q before prediction. If coders are uncertain, they record uncertainty separately in a notes field; they do not encode a range into the model. The predicted primary governance mode is generated only after point scores are assigned.

This matters because ranges can change the result. A Terraform provider with A=2, I=3, G=3, R=2 may be directly admitted. The same provider with A=3 and R=2 has a revocation deficit and requires policy or revocation controls before admission is defensible. The model is not indeterminate; the use context was underspecified. Point scoring forces the analyst to specify the context before prediction.

\subsection{11.3 Outcome Coding Rubric}\label{outcome-coding-rubric}

The dependent variable is not ``secure'' or ``insecure.'' It is the \textbf{primary observed admission mode}.

Because governance modes can stack, each case must be coded for both primary and secondary modes. The primary mode is the lowest governance layer without which the artifact would not be allowed in the relevant use context. Secondary modes are supporting controls that improve custody but are not the binding admission condition.

\begingroup\small
\begin{longtable}[]{@{}
  >{\raggedleft\arraybackslash}p{(\columnwidth - 4\tabcolsep) * \real{0.4000}}
  >{\raggedright\arraybackslash}p{(\columnwidth - 4\tabcolsep) * \real{0.3000}}
  >{\raggedright\arraybackslash}p{(\columnwidth - 4\tabcolsep) * \real{0.3000}}@{}}
\toprule\noalign{}
\begin{minipage}[b]{\linewidth}\raggedleft
Code
\end{minipage} & \begin{minipage}[b]{\linewidth}\raggedright
Outcome
\end{minipage} & \begin{minipage}[b]{\linewidth}\raggedright
Primary-mode coding rule
\end{minipage} \\
\midrule\noalign{}
\endhead
\bottomrule\noalign{}
\endlastfoot
0 & Admitted & Artifact is allowed in the relevant use context without special intermediary governance beyond ordinary baseline controls \\
1 & Proxied & Artifact is allowed only through an internal feed, mirror, remote repository, proxy, curated registry, or private marketplace \\
2 & Policy-mediated & Artifact is allowed only if it satisfies explicit policy constraints such as pinning, signing, allowlisting, permission reduction, or review \\
3 & Vendor-mediated & Artifact is allowed only through a commercial/cloud/platform wrapper that supplies catalog, scanning, identity, logging, support, isolation, deployment, or contractual controls \\
4 & Internalized & Artifact's function is allowed only after copying, wrapping, republishing, forking, or converting it into an institution-owned artifact \\
5 & Quarantined/rejected & Artifact is limited to experimentation, sandbox use, non-production paths, or blocked outright \\
\end{longtable}
\endgroup

A case can have multiple controls. The coder must still assign a primary mode. The primary mode is determined by the first binding gate in the observed policy or practice. If a public package is scanned and pinned but may only be consumed through an internal feed, the primary mode is proxied. If a GitHub Action is allowed from the public internet only when pinned to a full SHA and allowlisted, the primary mode is policy-mediated. If a public model is allowed only inside a cloud model catalog or enterprise hub, the primary mode is vendor-mediated. If a public Terraform module is copied into an internal module registry before use, the primary mode is internalized.

The prediction protocol is strict. The model must emit a primary predicted mode before the observed mode is scored. A prediction counts as a hit only if the predicted primary mode matches the independently coded primary observed mode. Secondary controls may be reported but cannot be used post hoc to rescue a failed primary prediction.

\subsection{11.4 Deterministic Prediction Function}\label{deterministic-prediction-function}

Predicted governance mode should be generated mechanically, not assigned by coder judgment.

Coders score only the inputs: A, I\_pre, G\_pre, R\_pre, S, Q, external closure availability, and route-change versus in-place-policy closure where relevant. A deterministic spreadsheet formula or script then applies:

\begin{enumerate}
\def\labelenumi{\arabic{enumi}.}
\tightlist
\item
  the closure rule;
\item
  the co-binding cascade rule;
\item
  the specificity rule;
\item
  the external-closure condition;
\item
  the route-change versus in-place-policy convention.
\end{enumerate}

The script emits one primary predicted governance mode. This prevents coders from unconsciously scoring toward the outcome they expect. Reliability is measured on the input scores. Predictive validity is measured on whether the fixed mapping matches the independently coded observed mode.

\subsection{11.5 Independence of Inputs and Observed Outcomes}\label{independence-of-inputs-and-observed-outcomes}

The main threat to H2 is partial circularity. If I/G/R scores and observed governance mode are both read from the same control document, the study may simply recode one artifact twice: first as a set of controls, then as a governance outcome.

The empirical protocol should therefore separate the input evidence from the outcome evidence as much as public data allows.

Input scores should be coded from control surfaces: identity mechanisms, ingress paths, revocation mechanisms, authority context, specificity, and scrutiny proxies. Observed primary mode should be coded from behavioral or gate-level evidence: what is actually allowed, blocked, required, or mediated at the admission point. Examples include whether a policy requires use of an internal feed, whether public actions are disabled unless allowlisted, whether extensions can be installed only if approved, whether models can be deployed only through a managed catalog, or whether MCP tools must pass through a gateway.

Where the same document contains both input controls and outcome rules, coders should mark the observation as same-source and report a robustness check excluding or down-weighting same-source observations.

Observed-mode coding should be performed blind to the generated prediction whenever possible.

\subsection{11.6 Scrutiny-Stratified Validation}\label{scrutiny-stratified-validation}

The validation design must test the positive claim without allowing scrutiny to become an after-the-fact explanation.

Before coding outcomes, each observation receives a scrutiny score Q using observable proxies: regulated sector, SOC 2 or ISO 27001 posture, FedRAMP or equivalent government authorization, formal procurement or customer-assurance obligations, security questionnaire exposure, public incident history, cyber-insurance review, or mission-critical workload status.

The main cross-sectional test is:

\textbf{For the same artifact/use-context class and execution-authority level, higher-Q observations should show stronger realized custody controls or more restrictive primary governance modes.}

This does not prove longitudinal convergence. It tests the observable implication of the convergence claim in cross-section. A longitudinal study would require repeated observations of the same institution or artifact class before and after increased scrutiny, such as a breach, audit finding, regulated-customer acquisition, FedRAMP pursuit, or enterprise procurement escalation.

This preserves falsifiability. A high-Q institution that routinely admits high-authority artifacts with weak identity, uncontrolled ingress, and weak revocation is a serious counterexample, not something to be dismissed after the fact.

\subsection{11.7 Base-Rate and Class-Balanced Evaluation}\label{base-rate-and-class-balanced-evaluation}

Raw hit rate is insufficient because the corpus may be class-imbalanced. Public evidence is likely to overrepresent proxied and policy-mediated modes, while internalized, quarantined, and rejected modes may be rarer or less publicly documented.

The validation should therefore report:

\begin{itemize}
\tightlist
\item
  majority-class baseline accuracy;
\item
  balanced accuracy;
\item
  per-class precision and recall;
\item
  macro-F1;
\item
  confusion matrix for primary predicted versus primary observed mode;
\item
  hit rate by binding-deficit class.
\end{itemize}

H2 is meaningful only if the model beats the majority-class baseline and performs nontrivially on rare governance modes. A high raw hit rate that merely predicts the modal outcome is not evidence of explanatory power.

\subsection{11.8 Pilot Power and Claim Discipline}\label{pilot-power-and-claim-discipline}

A 40-60 unit pilot is not powered to establish a clean causal interaction across all artifact classes and governance modes. It should be framed as a proof-of-concept for the instrument, a reliability study for the rubrics, and an initial directional test of H1-H4.

H4 in particular should be interpreted cautiously. With multiple input variables, a scrutiny score, and an A x Q interaction, the pilot can indicate whether the expected direction appears; it cannot support strong claims about effect size or generalization. A powered follow-up would require a larger corpus, more institutions per scrutiny stratum, and repeated observations across artifact classes and use contexts.

H1 will be evaluated by examining whether higher execution authority is associated with stronger custody strength, $\min(I,G,R)$, and secondarily with more restrictive governance modes. H3 will be evaluated descriptively by comparing predicted and observed governance modes for the same artifact class across different authority contexts.

The pilot's strongest contribution is therefore methodological: it tests whether independent coders can score the instrument reliably and whether the deterministic mapping beats plausible base-rate baselines.

\subsection{11.9 Parsimony and Overfitting Risk}\label{parsimony-and-overfitting-risk}

The model now contains several variables: execution authority, object identity, ingress path, revocation capacity, local custody specificity, scrutiny, and external closure availability. That creates an obvious parsimony concern. A skeptical reader could ask whether the framework explains divergent outcomes or simply has enough moving parts to fit them after the fact.

The answer is the prediction protocol.

The variables are not free-form explanations. They are scored ex ante using anchored rubrics. The closure rule is fixed. The tie-break rules are fixed. The dependent variable has an independent coding rubric. The model emits one primary predicted governance mode before the observed mode is scored. Secondary controls cannot rescue a miss. The six derivation cases are separated from held-out cases. Validation requires two coders and reported agreement.

The framework therefore trades minimalist elegance for operational falsifiability. It is more complex than a single-factor account because artifact admission is not governed by a single factor. But it is not unconstrained. Its discipline is not parsimony by variable count; it is parsimony by mechanism:

\textbf{authority determines required custody; weakest custody determines direct-admission failure; specificity and external closure determine governance response; scrutiny predicts where convergence appears first.}

A model with more variables than this would risk becoming descriptive bookkeeping. A model with fewer variables cannot distinguish Terraform providers from Terraform modules, package dependencies from CI actions, or model hubs from model catalogs without smuggling the missing distinctions back into prose.

\subsection{11.10 Replication Package}\label{replication-package}

The research submission should include a replication package.

At minimum, it should contain:

\begin{itemize}
\tightlist
\item
  the A/I/G/R/S/Q coding rubric;
\item
  the dependent-variable rubric;
\item
  the deterministic prediction script or spreadsheet;
\item
  the source corpus;
\item
  coded observations;
\item
  coder notes;
\item
  adjudication log;
\item
  reliability calculations;
\item
  confusion matrix and baseline comparisons.
\end{itemize}

The replication package is not cosmetic. It is what turns the Custody Envelope Threshold from an interpretive framework into a reusable instrument.

\section{12. Held-Out Evidence: MCP Servers and AI Tooling}\label{held-out-evidence-mcp-servers-and-ai-tooling}

This MCP section is current as of June 1, 2026. Because MCP governance surfaces are changing quickly, the claims here should be treated as a dated held-out evidence slice rather than a stable ecosystem survey.

MCP servers and AI tool integrations are the strongest held-out class because they were not part of the six-case derivation set but exhibit the same admission pressure.

MCP connects LLM applications to external data sources and tools. MCP servers can expose resources, prompts, and tools; tool calls may read data, invoke APIs, interact with repositories, query databases, inspect cloud state, or trigger workflows. That makes high-authority MCP servers artifact-admission objects, not merely configuration entries.

The framework predicts that arbitrary high-authority MCP server installation will not remain institutionally admissible in high-scrutiny environments. Many MCP servers score A=2 or A=3 because they expose enterprise data or action surfaces. In unmanaged form, their identity, ingress, and revocation can remain weak: public server lists, package-manager installs, GitHub repository references, mutable versions, broad tokens, ad hoc local configuration, and limited inventory.

The current evidence supports a narrower claim: \textbf{the governance surfaces the model predicts are already materializing.} It does not yet prove demand-side convergence across institutions.

First, the protocol itself is hardening. The 2025-06-18 MCP authorization specification required MCP servers to implement OAuth 2.0 Protected Resource Metadata and required clients to use it for authorization-server discovery. The 2025-11-25 authorization specification retained Protected Resource Metadata requirements, added Client ID Metadata Documents as a recommended client-registration path, allowed Dynamic Client Registration as a fallback, and required authorization servers to support OAuth Authorization Server Metadata or OpenID Connect Discovery mechanisms.

Second, registries and catalogs are appearing as discovery and admission surfaces. The official MCP registry describes itself as an app-store-like list of MCP servers, and Docker's MCP Catalog describes a curated collection of verified MCP servers packaged as Docker images and distributed through Docker Hub. Docker's MCP Catalog documentation also states that catalog servers run as isolated containers, and the Docker MCP Gateway includes server catalog management, secrets handling, OAuth integration, dynamic discovery, logging, and call tracing.

Third, gateway-mediated invocation is becoming an enterprise governance pattern. Microsoft Foundry's MCP governance documentation describes routing MCP traffic through an AI gateway that enforces authentication, rate limits, IP restrictions, audit logging, centralized observability, and public/private catalog reuse without modifying MCP servers or agent code.

Fourth, Cloudflare's MCP server portals place multiple MCP servers behind a single endpoint, allow administrators to customize which tools and prompts are exposed, and support Access-based authentication. Cloudflare's MCP portal logs include fields such as method, upstream server URL, session ID, tool-call name, success status, and authenticated user email. Cloudflare also documents routing MCP portal traffic through Cloudflare Gateway for HTTP logging and DLP scanning, including policies that can detect and block sensitive data sent to upstream MCP servers.

The evidence is not one-sided. Cloudflare explicitly discusses ``Shadow MCP'' discovery - detecting unauthorized remote MCP servers not accessed through a portal - which confirms both sides of the model: unmanaged MCP use is real, and higher-governance actors are building controls to find and mediate it. Cloudflare's reference architecture frames MCP server portals as a way to centralize discovery, logging, policy enforcement, DLP guardrails, user/tool access control, and authorized employee access to internal and third-party MCP servers.

The corrected claim is therefore:

\textbf{MCP is early held-out evidence that the predicted governance surfaces are materializing. It is not yet evidence that institutions have converged.}

The stronger empirical test is H4: high-scrutiny institutions should move first toward catalogs, gateways, OAuth/scoped authorization, tool allowlists, logging, DLP, and revocation paths.

\section{13. Uniform Point-Score Demonstration}\label{uniform-point-score-demonstration}

The following point-score demonstration uses representative high-scrutiny enterprise use contexts. It is not a completed validation corpus. It demonstrates how the coding protocol emits one primary predicted mode per unit when point scores are assigned.

\begin{enumerate}
\def\labelenumi{\arabic{enumi}.}
\item
  \textbf{Public package-registry dependency in production build/runtime, before institutional feed controls.} A=2, I=2, G=1, R=1, S=1, Q=2. The co-binding deficit is ingress plus revocation. The cascade closure is an internal feed or proxy that prevents future uncontrolled pulls and creates inventory. The primary predicted mode is \textbf{proxied}. Secondary controls include lockfiles, signatures, trusted publishing, and scanning.
\item
  \textbf{Third-party GitHub Action in CI with repository secrets, referenced by mutable tag.} A=3, I=1, G=1, R=1, S=1, Q=2. Identity, ingress, and revocation co-bind. Full-SHA pinning, allowlists, and permission constraints close the deficits inside the GitHub control plane. The primary predicted mode is \textbf{policy-mediated}. Secondary controls include least privilege, reviewed actions, and organization policy.
\item
  \textbf{Arbitrary public container image used as production base image.} A=3, I=2, G=1, R=1, S=1, Q=2. Ingress and revocation co-bind. Closing ingress requires a governed route: registry, mirror, proxy, or curated catalog. The primary predicted mode is \textbf{proxied}. Secondary controls include curated base lines, digest pinning, scanning, and image admission policy.
\item
  \textbf{Terraform provider used in controlled infrastructure automation with workspace-scoped credentials.} A=2, I=3, G=3, R=2, S=1, Q=2. There is no custody deficit: $\min(I,G,R)=2$ and A=2. The primary predicted mode is \textbf{admitted}. Secondary controls include signed providers, lockfiles, and mirrors.
\item
  \textbf{Public Terraform module used for production network/IAM design.} A=2, I=1, G=1, R=1, S=3, Q=2. Identity, ingress, and revocation co-bind, and local custody specificity is high. Safe use requires local platform knowledge. The primary predicted mode is \textbf{internalized}. Secondary controls include private module registries, golden modules, and platform review.
\item
  \textbf{Public VS Code extension on managed developer endpoint with source-code access.} A=2, I=2, G=1, R=1, S=1, Q=2. Ingress and revocation co-bind. The existing marketplace or endpoint route can be retained and constrained by policy. The primary predicted mode is \textbf{policy-mediated}. Secondary controls include allowlists, publisher constraints, version constraints, and private marketplaces when the route itself changes.
\item
  \textbf{Open model artifact requiring remote code in production serving.} A=3, I=1, G=1, R=1, S=2, Q=3. Identity, ingress, and revocation co-bind. An external model catalog or managed platform supplies multi-deficit closure. The primary predicted mode is \textbf{vendor-mediated}. Secondary controls include Safetensors, scanning, managed serving, and isolation.
\item
  \textbf{High-authority MCP server with write access to repositories or cloud APIs.} A=3, I=1, G=1, R=1, S=2, Q=3. Identity, ingress, and revocation co-bind. A gateway or catalog supplies identity, authorization, logging, and revocation surfaces. The primary predicted mode is \textbf{vendor-mediated}. Secondary controls include OAuth/scoped authorization, DLP, and tool-call logs.
\end{enumerate}

The demonstration is deliberately conditional. A different use context can change the score. A Terraform provider with organization-wide administrative credentials may score A=3 and require stronger revocation controls. A read-only local MCP server may score A=1 and be admitted with light guidance. A container image used only in a disposable sandbox may score A=1 and not require production-grade custody.

The coded validation must therefore evaluate \textbf{artifact x use context x institution}, not artifact label alone.

\section{14. Discussion: From Admission Rule to Governance Mode}\label{discussion-from-admission-rule-to-governance-mode}

The Custody Envelope Threshold changes the unit of analysis from ``open-source approval'' to \textbf{artifact-class admission}.

A single third-party-software policy cannot govern all external artifacts because artifact classes delegate different authority and require different custody closure. The institutional control surface for a package dependency is not the same as for a CI action, base image, Terraform module, model artifact, or MCP server. The admission question must therefore be artifact-class and context-specific: what authority does this artifact receive, what custody conditions bind it, what closure burden is acceptable, and which governance mode minimizes that burden?

The governance-economics implication is central. Institutions do not simply choose between trust and distrust. They choose among governance modes. Direct admission is available only when custody closure is sufficient. Proxying is selected when ingress is the cheapest binding deficit to solve. Policy mediation is selected when identity or permission control is cheaper than changing the artifact supply route. Vendor mediation is selected when an external platform can close multiple deficits more cheaply than the institution can. Internalization is selected when local specificity makes external custody insufficient. Quarantine or rejection is selected when closure burden exceeds the artifact's institutional value or risk tolerance.

This discussion is the bridge to the practitioner derivative. The research version should keep the governance-mode logic. The practitioner version can translate it into separate guidance for platform teams, security leaders, procurement, vendors, foundations, and investors.

\section{15. Predictions}\label{predictions}

The framework predicts several durable shifts.

First, direct public fetching will decline in serious production paths even as public artifact usage increases. More public dependency usage should produce more proxies, remote repositories, internal feeds, curated registries, and admission-aware catalogs, not less.

Second, CI/CD marketplace actions will be governed more strictly than language dependencies. Full-SHA pinning, allowlists, restricted permissions, and reviewed action catalogs should become ordinary in high-value repositories because CI actions operate at the software production boundary.

Third, container base images will consolidate around curated lineages and controlled ingress. Enterprises will prefer official, verified, hardened, internally rebuilt, or cloud-provider-mediated base images over arbitrary public images. Runtime admission policies should increasingly enforce this distinction.

Fourth, Terraform providers will remain more admissible than their raw authority might suggest because signing, lockfiles, checksums, and mirrors provide a strong custody envelope. Public modules will continue moving toward internal registries, wrappers, and golden modules because their binding risk is architectural rather than merely distributional.

Fifth, developer-environment extensions will become governed infrastructure artifacts. AI agents, terminal-capable extensions, cloud-connected IDEs, and local automation tools will make unmanaged extension ecosystems unacceptable in high-assurance environments.

Sixth, open model hubs will split into discovery layers and admission layers. Raw hubs will remain central to research and experimentation. Production use will flow through model catalogs, managed endpoints, enterprise hubs, scanner-backed registries, and internal approval systems that close the custody envelope.

The MCP prediction adds a seventh: high-authority AI tool servers will move toward catalogs, gateways, scoped permissions, identity pinning, and audit-mediated invocation. Raw arbitrary tool installation will remain acceptable in experimentation but become increasingly inadmissible in production AI-agent environments.

\section{16. Threats to Validity}\label{threats-to-validity}

\textbf{Construct validity.} The 0-3 rubric simplifies complex controls. Revocation capacity, for example, compresses inventory, block path, rebuild automation, credential response, and audit proof into one score. The rubric is useful only if anchors remain explicit and coders can apply them consistently.

\textbf{Internal validity.} The six derivation cases were selected because they display visible divergence. That improves explanatory contrast but creates selection risk. A held-out validation set is required before making stronger empirical claims.

\textbf{External validity.} Public documentation reveals platform features and recommended controls, not necessarily actual internal procurement decisions. Institutional behavior is partly inferred from observable control surfaces. A stronger study would use internal policy documents, interviews, or anonymized enterprise admission records.

\textbf{Reliability.} Without independent coders, the instrument remains analyst-scored. Reporting inter-rater reliability is necessary for peer-reviewed validation.

\textbf{Temporal validity.} Artifact ecosystems change quickly. New registry features, signing standards, cloud catalogs, AI tool protocols, and enterprise policy systems can move an artifact class from weak custody to strong custody. The framework should be applied to current artifact/use-context pairs, not frozen historical labels.

\textbf{Scope validity.} The framework intentionally excludes legal, semantic, export-control, behavioral, and business-purpose adequacy. Those gates can reject an artifact after technical custody closes. Treating them as out of scope preserves falsifiability but limits the framework's domain.

\section{17. What Would Falsify or Weaken the Framework}\label{what-would-falsify-or-weaken-the-framework}

The framework would be weakened by evidence that high-scrutiny institutions routinely admit high-authority external artifacts with weak custody closure and no mediating governance mode. Examples would include mutable third-party CI actions with secrets, arbitrary public container images in production clusters, unmanaged editor extensions on sensitive developer endpoints, raw model artifacts requiring remote code in regulated production serving, or arbitrary MCP servers with write access to enterprise systems - all admitted without pinning, proxying, cataloging, gateway mediation, inventory, logging, or revocation paths.

The framework would also be weakened if strong custody closure had little relationship to admission outcomes. If artifacts with pinned identity, controlled ingress, bounded authority, and rapid revocation were routinely rejected for technical-institutional custody reasons internal to the framework, then the model would be overstating custody closure and missing a stronger determinant.

The model is not weakened when an artifact is rejected for legal, semantic, safety, privacy, export-control, or business-purpose reasons after custody is closed. Those are separate gates. They may be decisive, but they are outside the Custody Envelope Threshold.

\section{18. From Model to Research Program}\label{from-model-to-research-program}

The next artifact is an \textbf{artifact-class admission profile}: a minimum admissibility profile for each artifact class and use context.

Each profile should specify:

\begin{itemize}
\tightlist
\item
  authority score;
\item
  identity requirements;
\item
  allowed ingress routes;
\item
  revocation requirements;
\item
  local custody specificity;
\item
  permitted governance modes;
\item
  logging and audit evidence;
\item
  owner roles;
\item
  review triggers;
\item
  downstream legal, semantic, safety, privacy, export, or business-purpose gates.
\end{itemize}

The research version should test whether real institutions approximate these profiles in public and internal policies. The practitioner version should convert them into operating checklists.

The central claim remains unchanged but sharper: institutions do not admit external artifacts because they are useful. They admit them when usefulness can be converted into authority-scaled custody at acceptable governance burden.

\section{Data and Code Availability}\label{data-and-code-availability}

The preregistered protocol, coding rubrics, deterministic prediction function, source-corpus templates, coded-data templates, analysis plan, and replication-package materials are available on OSF at \url{https://doi.org/10.17605/OSF.IO/E57FJ}. The associated OSF project is available at \url{https://osf.io/kc3ue/}. No pilot observations had been coded at the time of preregistration.

\section{References}\label{references}

{[}1{]} Supply-chain Levels for Software Artifacts (SLSA), ``SLSA: Supply-chain Levels for Software Artifacts,'' \url{https://slsa.dev/}.

{[}2{]} National Institute of Standards and Technology, ``Secure Software Development Framework (SSDF) Version 1.1: Recommendations for Mitigating the Risk of Software Vulnerabilities,'' NIST SP 800-218, 2022, \url{https://csrc.nist.gov/pubs/sp/800/218/final}.

{[}3{]} OpenSSF, ``Secure Supply Chain Consumption Framework (S2C2F),'' \url{https://github.com/ossf/s2c2f}.

{[}4{]} J. P. Anderson, ``Computer Security Technology Planning Study,'' ESD-TR-73-51, 1972, \url{https://apps.dtic.mil/sti/tr/pdf/AD0758206.pdf}.

{[}5{]} J. H. Saltzer and M. D. Schroeder, ``The Protection of Information in Computer Systems,'' Proceedings of the IEEE, 1975.

{[}6{]} O. E. Williamson, ``Transaction-Cost Economics: The Governance of Contractual Relations,'' Journal of Law and Economics, 1979.

{[}7{]} M. Ohm, H. Plate, A. Sykosch, and M. Meier, ``Backstabber's Knife Collection: A Review of Open Source Software Supply Chain Attacks,'' DIMVA, 2020.

{[}8{]} N. Zahan, T. Zimmermann, P. Godefroid, B. Murphy, C. Maddila, and L. Williams, ``What are Weak Links in the npm Supply Chain?'' ICSE-SEIP, 2022.

{[}9{]} A. Birsan, ``Dependency Confusion: How I Hacked Into Apple, Microsoft and Dozens of Other Companies,'' 2021, \url{https://medium.com/@alex.birsan/dependency-confusion-4a5d60fec610}.

{[}10{]} P. Ladisa, H. Plate, M. Martinez, and O. Barais, ``SoK: Taxonomy of Attacks on Open-Source Software Supply Chains,'' IEEE Symposium on Security and Privacy, 2023.

{[}11{]} OpenSSF, ``npm Best Practices Guide,'' \url{https://github.com/ossf/package-manager-best-practices/blob/main/published/npm.md}.

{[}12{]} Google Cloud, ``Artifact Registry remote repositories,'' \url{https://docs.cloud.google.com/artifact-registry/docs/repositories/remote-overview}.

{[}13{]} Microsoft Learn, ``What are upstream sources? --- Azure Artifacts,'' \url{https://learn.microsoft.com/en-us/azure/devops/artifacts/concepts/upstream-sources}.

{[}14{]} npm Docs, ``About registry signatures,'' \url{https://docs.npmjs.com/about-registry-signatures/}.

{[}15{]} npm Docs, ``Trusted publishing,'' \url{https://docs.npmjs.com/trusted-publishers/}.

{[}16{]} GitHub Docs, ``Secure use reference,'' \url{https://docs.github.com/en/actions/reference/security/secure-use}.

{[}17{]} GitHub Advisory Database, ``tj-actions/changed-files compromised,'' GHSA-mrrh-fwg8-r2c3, \url{https://github.com/advisories/GHSA-mrrh-fwg8-r2c3}.

{[}18{]} Docker Docs, ``Docker Hub trusted content,'' \url{https://docs.docker.com/docker-hub/image-library/trusted-content/}.

{[}19{]} Docker Library, ``Docker Official Images,'' \url{https://github.com/docker-library/official-images}.

{[}20{]} Sigstore Docs, ``Policy Controller Overview,'' \url{https://docs.sigstore.dev/policy-controller/overview/}.

{[}21{]} HashiCorp Developer, ``Terraform provider signing,'' \url{https://developer.hashicorp.com/terraform/cli/plugins/signing}.

{[}22{]} HashiCorp Developer, ``Terraform dependency lock file,'' \url{https://developer.hashicorp.com/terraform/language/files/dependency-lock}.

{[}23{]} HashiCorp Developer, ``terraform providers mirror,'' \url{https://developer.hashicorp.com/terraform/cli/commands/providers/mirror}.

{[}24{]} HashiCorp Developer, ``Terraform module sources,'' \url{https://developer.hashicorp.com/terraform/language/modules/configuration}.

{[}25{]} Visual Studio Code Docs, ``Workspace Trust,'' \url{https://code.visualstudio.com/docs/editing/workspaces/workspace-trust}.

{[}26{]} Visual Studio Code Docs, ``Enterprise support for extensions,'' \url{https://code.visualstudio.com/docs/enterprise/extensions}.

{[}27{]} Microsoft Developer Blog, ``Security and Trust in Visual Studio Marketplace,'' \url{https://developer.microsoft.com/blog/security-and-trust-in-visual-studio-marketplace}.

{[}28{]} Hugging Face Docs, ``Malware scanning,'' \url{https://huggingface.co/docs/hub/en/security-malware}.

{[}29{]} Hugging Face Docs, ``Pickle scanning,'' \url{https://huggingface.co/docs/hub/en/security-pickle}.

{[}30{]} Hugging Face Transformers, ``Security policy,'' \url{https://github.com/huggingface/transformers/security}.

{[}31{]} Microsoft Learn, ``Microsoft Foundry Models overview,'' \url{https://learn.microsoft.com/en-us/azure/foundry/concepts/foundry-models-overview}.

{[}32{]} Hugging Face Docs, ``Hugging Face on Azure - Security,'' \url{https://huggingface.co/docs/microsoft-azure/en/security}.

{[}33{]} Model Context Protocol, ``MCP Specification, 2025-06-18: Authorization (OAuth 2.0 Protected Resource Metadata, RFC 9728),'' \url{https://modelcontextprotocol.io/specification/2025-06-18/basic/authorization}.

{[}34{]} Model Context Protocol, ``MCP Specification, 2025-11-25: Authorization,'' \url{https://modelcontextprotocol.io/specification/2025-11-25/basic/authorization}.

{[}35{]} Model Context Protocol, ``Registry,'' \url{https://github.com/modelcontextprotocol/registry}.

{[}36{]} Docker Docs, ``Docker MCP Catalog and Toolkit,'' \url{https://docs.docker.com/ai/mcp-catalog-and-toolkit/}.

{[}37{]} Microsoft Learn, ``Govern MCP tools with Azure AI Foundry Agent Service,'' \url{https://learn.microsoft.com/en-us/azure/ai-foundry/agents/how-to/tools/governance}.

{[}38{]} Cloudflare Docs, ``MCP Server Portals,'' \url{https://developers.cloudflare.com/cloudflare-one/access-controls/ai-controls/mcp-portals/}.

{[}39{]} Cloudflare Blog, ``Introducing MCP server portals,'' \url{https://blog.cloudflare.com/enterprise-mcp/}.

{[}40{]} Open Science Framework, ``Custody Envelope Threshold: Artifact Admission Pilot and Replication Package,'' DOI 10.17605/OSF.IO/E57FJ, \url{https://doi.org/10.17605/OSF.IO/E57FJ}.

\end{document}